\documentstyle[12pt]{article}
\begin{document}
\begin{flushright}
PRA-HEP-92/14\\
\today
\end{flushright}
\vspace{5ex}
\begin{center}
{\LARGE \bf
Parton-hadron duality in event generators}\\[0.8in]
{\sc J.Ch\'{y}la} and {\sc J.Rame\v{s}}\\[0.1in]
{\it Institute of Physics, Czechoslovak Academy of Sciences}\\
{\it Prague, Czechoslovakia}\footnote{Postal address: Na Slovance 2,
180~40~~Prague 8, Czechoslovakia}\\
\vspace{.8in}
{\bf Abstract}\\
\end{center}
\begin{quotation}
The validity of local parton-hadron duality
within the framework of HERWIG and JETSET event generators is investigated.
We concentrate on ${\rm e}^{+}{\rm e}^{-}$ annihilations in LEP 2
energy range as these interactions provide theoretically the
cleanest condition for the discussion of this concept.
\end{quotation}
\renewcommand\thefootnote{\arabic{footnote}}
\newpage
\section{Introduction}
The concept of parton hadron duality (PHD) and in particular its local version
attempts to answer the question of the relation between the properties of
experimentally observed hadrons and the assumed underlying parton dynamics.
A few years ago the St.Petersburg school has gone beyond the original global
version of the duality idea and has argued in favour of much closer relation
between the single particle inclusive spectra of partons and hadrons
\cite{Azimov}.  They have developed powerful theoretical tools to calculate
within perturbative QCD partonic spectra in great detail, taking into account
various subtle effects (for comprehesive review of this topic see, for
instance, \cite{book}). In converting their results into the statements
concerning hadrons they, however, crucially rely on two important
 assumptions.
First, the independent fragmenation model is used to hadronize the partonic
configurations originating from perturbative cascading. Secondly, partonic
cascades are allowed to evolve down to rather small (timelike) virtualities
of the order of pion mass. As neither of these assumptions is
incorporated in the currently widely used generators based on either the
string or cluster fragmentation, we have undertaken a detailed
investigation of the PHD within two distinct event generators which
successfully describe vast amount of experimental data from various
collisions. For related work see also \cite{L1,L2}.
Both HERWIG (we use its 5.3 version) and JETSET (version 7.2) use in their
respective hadronization stages algorithms which do not allow a sensible,
 i.e. reasonably unambiguous interpretation of produced hadrons as being
fragments of a particular single parton. In our study we have addressed two
closely related questions:
\begin{itemize}
\item to what extent do the hadronic distributions reflect those
   of the perturbatively produced partons
\item how much do the partonic spectra in HERWIG and JETSET differ from each
 other
\end{itemize}
In the local PHD picture the first question has a very simple answer, at
 least
if the fragmentation function advocated by the authors of \cite{Azimov} is
employed: hadronic spectra are proportional to those of the partons with the
proportionality factor of the order of unity. This is not the case in either
HERWIG or JETSET generator. These generators are both based on the partonic
cascade as described by perturbative QCD so that the starting point is in
principle the same as in \cite{Azimov}. The partonic cascades in HERWIG and
JETSET are, however, neither identical nor fully equivalent to the analytic
calculations of \cite{Azimov} so that differences do appear already on the
parton level. Moreover, the influence of the hadronization stage turns out to
be very important in its effects on the  hadronic spectra. What we basicaly
observe is that quite different configurations of partons yield much the same
hadronic spectra. Indeed while HERWIG and JETSET are quite different
as far as partonic spectra are concerned they  yield remarkably similar
results for hadrons. This indicates that the interplay between the
perturbative and
hadronization stages in hard scattering processes is very important,
nontrivial and model dependent. Because of this ambiguity in the relation
between the partonic and hadronic characteristics the concept of local PHD
looses much of its intuitive appeal and predictive power.

In order to investigate these questions in relatively ''clean''
conditions, we have concentrated on the ${\rm e}^{+}{\rm e}^{-}$ annihilations
into hadrons at 200 GeV center of mass energy, i.e.\ on LEP 2
energy range. There the perturbative cascades, though model dependent,
are already rather well developed and so  the whole perturbation
theory machinery seems to be well justified and under control.

Any comparison between several sets of results depends on the quantities
selected for that purpose. We have chosen
the following ones:
\begin{itemize}
\item multiplicity distribution
\item single particle inclusive distributions in the standard variables
$z$ and $p_{t}^{2}$ where
\begin{equation}
z=\ln \left(\frac{1}{x_{p}}\right);\;\;\;\;
x_{p}=\frac{2p}{\sqrt{s}}
\label{definition}
\end{equation}
where $p_{t}$ denotes the transversal
component of particle momenta with respect to the thrust axis.
\item factorial moments of particle multiplicities (intermittency measure)
\end{itemize}
The first two types of them are those discussed in \cite{Azimov} while
the factorial moments of particle multiplicities in small regions of
phase space \cite{Bialas1,Bialas2} are commonly considered as a clear
manifestation of underlying partonic cascading \cite{Ochs}.

\section{A few remarks on generators}
We briefly recall some of the important features of HERWIG and JETSET
generators and in particular those of their parameters which will play an
essential role in the following discussion of local PHD. In HERWIG the event
simulation proceeds in four stages:
\begin{enumerate}
\item hard scattering, i.e. in our case
 ${\rm e}^{+}{\rm e}^{-}\rightarrow {\rm q}\overline{\rm q}$
\item evolution of parton showers on all partonic legs, here
q$\overline{\rm q}$
\item formation of colorless clusters out of partonic final state
 ${\rm q}\overline{\rm q}$ pairs
\item decay of these clusters into hadrons
\end{enumerate}
In JETSET the first two steps are in principle similar although the details
of shower evolution differ from HERWIG. The main difference between these
generators concerns, however, the hadronization stage. Instead of the
formation of colorless clusters \mbox{JETSET}  spans relativistic string on
the products of partonic cascade which then breaks into observable hadrons.

There are many parameters which determine the details of each of these steps,
but the following ones are essential for fixing the relative importance
and interplay between the parton shower and hadronization stages in HERWIG:
\begin{itemize}
\item $QCDLAM$: the usual QCD $\Lambda$-parameter
\item $VQCUT,VGCUT$: parameters setting, when added to parton masses,
 the minimal parton virtuality in timelike cascades. The default options are
 such that for both quarks and gluons the  minimal timelike virtuality is
 about 0.9 GeV.
\item $CLMAX$: the decisive parameter for the hadronization stage of
 HERWIG. It forces the
 colorless clusters, produced in the perturbative cascade,
 with mass squared
  above $CLMAX^{2}+(m_{q}+m_{\bar{q}})^{2}$ to split, before
 hadronization, into lower mass ones. The default value is 3.5 GeV.
 Although in typical events the action of $CLMAX$ parameter is limitted
 it plays quite an essential role in defining the relative role of
 partonic cascades and hadronization stages. As will be shown below it is
 particularly important in low parton number events.
\end{itemize}
In JETSET analogous role is played by the parameters:
\begin{itemize}
\item PARJ(81): analogue of $QCDLAM$
\item PARJ(82): invariant mass cutoff on parton virtualities, similar in
 effect to $VGCUT$, $VQCUT$ in HERWIG.
\item there is no direct analogue of $CLMAX$ parameter.
\end{itemize}
In JETSET the user
can choose betweeen the leading log parton showers and exact fixed order
matrix element approaches. In HERWIG parton showers cannot be
straightforwardly switched off and they always accompany the hard scattering
subprocess, be it ${\rm e}^{+}{\rm e}^{-}\rightarrow {\rm q}\overline{\rm q}$
or  ${\rm e}^{+}{\rm e}^{-}\rightarrow {\rm q}\overline{\rm q}g$.
Nevertheless to study the
relation between partons and hadrons in theoretically cleanest conditions we
have concentrated in both generators on the
${\rm e}^{+}{\rm e}^{-}\rightarrow{\rm q}\overline{\rm q}$ subprocess and
selected the parton shower option in JETSET. In our quest to understand the
relation between the parton and hadron spectra we have
\begin{itemize}
\item compared the hadron as well as parton characteristics, specified
 above, for the currently ''best'' sets of parameters of HERWIG
 and JETSET
\item compared parton distributions with the corresponding hadronic ones, in
 HERWIG as well as in JETSET
\item looked separately on events characterized by different number of
 partons. ''Tagging'' on this parton number allows us to see
 clearly under what circumstances are the hadronic
 characteristics direct manifestations of the underlying partonic ones.
\item varied the values of $VQCUT,VGCUT,CLMAX$ and looked at the changes in
 the relation between partonic and hadronic spectra
\end{itemize}

\section{Discussion of the results}
The basic results of extensive simulations with HERWIG and JETSET at LEP2
energy are presented in Fig.1-7. We stress that neither generator has been
tuned especially for our purposes and we have taken the currently ''best''
sets of their respective parameters.
\subsection{Multiplicity distributions}
Fig.1 displays the comparison between the partonic as well as hadronic
multiplicity distributions in HERWIG and JETSET. Already this simplest
quantity signals the basic message:
while for the hadrons these models predict results which are rather
close to each other (typically within 10-15 \%) they differ vastly on
the level of partons!
Despite the large difference in the average number of perturbatively
produced partons (9 in HERWIG vs. 15.5 in JETSET)
the hadronic multiplicity distributions
are much closer in shape as well in average values: 42.5 in
 HERWIG vs. 48 for JETSET. This large difference between the average
parton and hadron multiplicities is in sharp contrast with the
results of \cite{Azimov}, where the number of partons is much closer
to the number of hadrons.

\subsection{Single particle inclusive spectra}
The same message can be read off  Fig.2, where the single particle
spectra in $z$ and $p_{t}^{2}$, from both HERWIG and JETSET,
are compared on partonic as well as hadronic levels.
 The pronounced difference in $z$
distributions of partons in the region close to zero
comes from the region of phase space where one of the partons carries
nearly all available momentum. The most important difference between HERWIG
and JETSET partonic distributions is,
however, observed for large $z$ and small $p_{t}^{2}$, i.e. for soft partons
which dominate the total multiplicity. We see that HERWIG $z$-distribution
is significantly lower than that of JETSET down to $z=2$ and practically
vanishes for $z>5$. Despite these dramatic differences on the level of
partons, the corresponding hadronic distributions are quite similar even in
the large $z$ region.
 For $p_{t}^{2}$ distributions of partons the dramatic
difference in the region close to $p_{t}^{2}=0$ is a direct manifestation of
the cut on minimal virtuality of partons as set by the parameters
$VQCUT,VGCUT$. The position of the maximum in HERWIG spectrum is in fact
proportional to them. In JETSET there does not seem to be an analogous
effect.
This difference between HERWIG and JETSET is, however, again not
reflected in the corresponding hadronic $p_{t}^{2}$ distributions, which look
practically indistinguishable. The
slightly higher hadronic multiplicity in JETSET is then reflected in somewhat
higher values of hadronic $z$ spectra in the region around $z=4$.

In Fig.3 we compare the partonic and hadronic spectra obtained with both
generators by plotting (as solid lines) the ratia
\begin{equation}
r(w)=\left (\frac{1}{N_{events}}\frac{dN}{dw_{partons}}\right )\left /
\left (\frac{1}{N_{events}}\right .
\frac{dN}{dw_{hadrons}}\right );\;\;\;w=z,p_{t}^{2}
\label{ratia}
\end{equation}
of appropriately normalized partonic over hadronic
distributions, which should be approximately constant if local PHD holds.
Clearly rather large deviations from constancy in most of the phase space and
for both $z$ and $p_{t}^{2}$ are observed, the pattern of these
violations being, except for $z$ close to the upper limit, similar in HERWIG
as in JETSET.
In both generators a large part of this effect can be traced back to the fact
that their partonic showers are stopped at much higher virtualities than
in \cite{Azimov}. This is demonstrated by dotted lines in Fig.3 which
correspond, for HERWIG as well as JETSET, to lower virtuality cut-off
$Q_{0}=0.2$ GeV.
For technical reasons this low virtuality cut-off requires, in both
generators, simultaneous lowering of the QCD $\Lambda$-parameter. In our case
we have taken it to be 0.04 GeV.

 The most dramatic effect occurs for the ratio $r(z)$
which for $Q_{0}=1.0$ GeV was rapidly decreasing function of $z$ in the whole
phase space, while now this ratio is nearly constant in the large interval
$z\in (1,6)$. Similar effect is observed for $p_{t}^{2}$ spectra, in
particular on low $p_{t}^{2}$ region.
There is thus no doubt that local PHD is much better
reproduced in both
HERWIG and JETSET for small virtuality cut-off $Q_{0}=0.2$ GeV.
However, neither HERWIG nor JETSET can accommodate such low values of $Q_{0}$
and still describe the available experimental data as accurately as with the
$Q_{0}$ in the region of 1 GeV.
In the regions of $x_{p}$ close to 0 and 1 the deviations from constancy are
very large even for $Q_{0}=0.2$ GeV but this is not
surprising as these are also the regions where the analytic calculation of
\cite{Azimov} contain subtle effects not included in the generators.
An interesting difference between HERWIG and JETSET $z$ distributions is
observed in large $z$ region, i.e. for $x_{p}\rightarrow 0$, which is
populated by soft partons, and where the ratia behave quite oppositely.
All this suggests that the validity of local PHD
relies heavily on evolving the partonic showers to rather small scales
comparable to the pion mass.

In order to understand the differences between HERWIG and JETSET in more
detail we have furthermore
subdivided all events in three classes
according to the number of perturbatively produced partons.
\begin{itemize}
\item small number of partons: 2-6
\item moderate number of partons: 7-11
\item large number of partons: more than 11
\end{itemize}
In Fig.4 the comparison between partonic $z$ and $p_{t}^{2}$ spectra
from HERWIG and JETSET is done for each of the classes separately. We see
that the differences increase with decreasing number of underlying partons.
For all three classes of events the corresponding hadronic
spectra (not displayed) are, however, again much closer as in the case of the
full samples.

We now come to the interplay between the parton shower and cluster
decay stages in HERWIG event generator. In Fig.5 we plot separately for
each of the three classes of events the
$z$ and $p_{t}^{2}$ distributions for partons and hadrons. In the case of
hadronic distributions we moreover show two sets of curves corresponding
to two different values of the $CLMAX$ parameter. Beside the default
value $CLMAX=3.5$ GeV we plot also the results corresponding to
$CLMAX=50$ GeV. This large value effectively means that we do not force
large mass clusters to split into smaller ones before hadronization
and thus come close to the original formulation of HERWIG. We conclude that
\begin{enumerate}
\item Small value of $CLMAX$ is more effective in events with small
 number of partons. For them perturbative branching itself leads
 on average to small number of heavy clusters. Once small $CLMAX$ is
 taken, all these heavy clusters are first split into smaller
 ones and only then allowed to hadronize. After this step the cluster
 mass distributions are essentially the same
 irrespective of the original parton number. The only trace of the
 originally different parton numbers then remains in the smaller
 cluster multiplicity.
\item The differences between the shapes as well as magnitudes of
 parton and hadron spectra are large in the whole phase space
 and depend sensitively on the number of partons.
\item The differences between the two hadronic spectra, corresponding
 to $CLMAX=3.5$ GeV and $CLMAX=50$ GeV show that for events with
 small parton numbers the initial part of the hadronization stage
 i.e.\ the cluster splitting is crucially important. This shows
 why and how the hadronization  stage may
 significantly influence the relation between the parton and hadron
 spectra.
\end{enumerate}

Taken together the message contained in the preceding observations is
simple and clear: While on the level of hadrons different models
give very similar results, these results can originate
from very different partonic distributions. Moreover, the effects and
importance of the hadronization stage are nontrivial and do depend on
particular partonic configuration.
All this demonstrates that the concept of local PHD does not hold within
either HERWIG
or JETSET event generators and is thus more ambiguous than as claimed in
\cite{Azimov}.

\subsection{Intermittency analysis}
The intermittency phenomenon, as quantified by the factorial moments
in small phase space regions \cite{Bialas1,Bialas2}
is commonly regarded as a convincing
evidence for the underlying partonic cascading process with the
selfsimilarity property \cite{Ochs}. As such it should also provide
the evidence
for the local PHD. To check this assumption we have carried out
several simple tests using both HERWIG and JETSET event generators.
The question we ask ourselves is similar
as before: to what extent is the
observed intermittency behaviour on the level of hadrons a direct
consequence of the underlying partonic cascade?

To find the answer we have calculated the conventional factorial moments of
the $i-th$ rank in two dimensions (rapidity versus azimuthal angle)
\begin{equation}
F_{i}(y,\phi)=\frac{1}{N_{events}}\sum_{events}
\frac{\sum_{k=1}^{n_{bins}}\langle n_{k}(n_{k}-1)
\cdots (n_{k}-i+1)\rangle }{(\langle n\rangle /n_{bins})^{i}}
\label{fi}
\end{equation}
where $\langle n\rangle$ is
 the average number of particles in the full $\Delta y-\Delta
\phi$ region accepted (we have taken full azimuthal coverage and
$y\in (-2,2)$)
and $n_{bins}$ denotes the number of bins in this
two-dimensional space, given as $4^{b}$ with $n_{div}=2^{b}$ defining the
number of
divisions in each of the two directions. We have constructed these moments
for partons as well as hadrons as functions of $b$ or $n_{div}$.
Then we have compared for $i=2,3,4,5$
\begin{itemize}
\item the results from HERWIG and JETSET (Fig.6a)
\item the results for partons with those of hadrons (Fig.6b)
\item the results corresponding to events with different
 number of partons (Fig.7)
\end{itemize}
On the basis of the plots displayed in Figs.6,7 we draw the
following conclusions, which point in the same direction as those
of the preceding paragraphs:
\begin{enumerate}
\item Despite the large differences on the partonic level
  HERWIG and JETSET give very similar results for the hadronic factorial
  moments (\ref{fi}). This conclusion holds well for the full sample of
  events as well as for all three classes of events defined above
(corresponding plots similar to Fig.6a are omitted), i.e. is independent of
the number of underlying partons.
\item For HERWIG we find a remarkable agreement between the partonic and
hadronic
 factorial moments in the region where the former show the appropriate
 rising behaviour. It is clear that this region can not be large as
 the number of partons is much smaller than that of the hadrons.
 In order to see the effect on the parton level clearly we have
therefore chosen finer
steps in the division of $\Delta$y-$\Delta\phi$ interval for partonic
moments. We have, however, seen that much the same behaviour of hadronic
factorial moments results even for the case where there are so little
partons that their own factorial moments vanish. So again we see that the
property of hadrons usually considered  as a direct consequence of the
presence of partonic cascade can equally well result from the effects of the
hadronization stage. In other words there is some kind of duality but of
different sort that originally proposed in \cite{Azimov}, namely the duality
between the partonic and hadronization stages within the whole event
generation.
\item Except for $F_{2}$ the factorial moments in HERWIG as well as JETSET
are practically independent of the number of partons in the event.
In Fig.7 we present results from JETSET, but the same picture is obtained for
HERWIG as well. For the purpose of this comparison we have somewhat
changed the definition of the three classes of events (2-10,11-18,19 and
more) which is more appropriate for JETSET due to its higher average parton
multiplicity.
 This at first sight surprising
observation again shows that there is no simple relation between the partonic
and hadronic properties. The
fact that events with small number of partons give smaller  $F_{2}$ than
those with many partons may be a simple reflection of lower overall hadronic
multiplicity in these events. In HERWIG the approximate indepedence of
the factorial moments has a simple explanation in the interplay between the
partonic cascade and the hadronization stage. What happens is that in the
case of small parton number the action of the $CLMAX$ parameter creates out
of the small number of heavy clusters, produced by purely perturbative
cascade, much larger number of lighter ones. Instead of partonic cascade we
have ''cluster'' cascade, which, however, produces essentially the same
behaviour of hadronic factorial moments. In JETSET we can offer no such
simple explanation but the effect holds as well.
\end{enumerate}

\section{Summary and conclusions}
In this paper we have discussed the relation between partons and hadrons in
two different, widely used event generators, HERWIG and JETSET. We
concentrated on ${\rm e}^{+}{\rm e}^{-}$ annihilations at LEP 2 energies as
these conditions provide theoretically the cleanest place for investigation of
such a relation. Our simulations show that this relation is complicated and
generator dependent. The idea of local PHD as suggested in \cite{Azimov} is
not realized in either of these models. To large extent, this is due to the
fact that the analytical calculations of \cite{Azimov} rely in a crucial way
on the use of independent fragmenation model coupled with the assumption that
partonic showers are allowed to evolve down to virtualities of the order of
the pion mass, while both HERWIG and JETSET stop their respective cascades at
much higher virtualities.

On the other hand we have found evidence for the strong interplay between the
effects
of parton showers and those of the hadronization stage. Clear example of
such an interplay is provided by factorial moments in narrow phase space
region
where the intermittent behaviour originates equally from perturbative parton
branching as well as from colorless cluster splitting.
This indicates that we can not check details of partonic evolution without
the knowledge of hadronization mechanism and vice versa.

\newpage
\parindent 0.cm
{\bf Figure captions.}\\

\vspace {0.9cm}
Fig.1:
Multiplicity distributions of partons (a) and hadrons (b) from {\mbox HERWIG}
and {\mbox JETSET}.

\vspace {0.9cm}
Fig.2:
Single particle distributions in $z$ and $p_{t}^{2}$ of hadrons (a)
and partons (b) from {\mbox HERWIG} and {\mbox {JETSET}.

\vspace {0.9cm}
Fig.3:
Ratia $r(z)$, $r(p_{t}^{2})$ from {\mbox JETSET} (a) and {\mbox HERWIG} (b)
for two values of the lower virtuality cut-off $Q_{0}$.

\vspace {0.9cm}
Fig.4:
Single particle distributions of hadrons from HERWIG and {\mbox JETSET},
as functions of $z$ and $p_{t}^{2}$, for the three classes of events defined
in the text.

\vspace {0.9cm}
Fig.5:
Partonic and hadronic distributions in $z$ and $p_{t}^{2}$ for the three
classes of events defined in the text as obtained from HERWIG.
For the hadrons the spectra corresponding to two different values of $CLMAX$
parameter are displayed.

\vspace {0.9cm}
Fig.6:
Comparison of factorial moments $F_{i}(y,\phi)$
between HERWIG and JETSET (a), and between partons and hadrons within HERWIG
(b).

\vspace {0.9cm}
Fig.7:
Dependence of factorial moments $F_{i}(y,\phi)$ on the number of partons as
obtained from JETSET.
\end{document}